\documentclass[aps,prd,preprint,groupedaddress]{revtex4-1}

\usepackage{graphicx}
\usepackage[table,xcdraw]{xcolor}
\usepackage{xcolor}
\usepackage{amsmath}
\begin{document}

\title{ Configuration entropy and confinement/deconfinement transiton in holographic QCD}

\author{Nelson R. F. Braga\thanks{\href{mailto:nrfbraga@gmail.com}{nrfbraga@gmail.com}},\, O.~C.~Junqueira\thanks{\href{mailto:octaviojunqueira@gmail.com}{octaviojunqueira@gmail.com}} }
\affiliation{Universidade Federal do Rio de Janeiro---Instituto de F\'isica, Caixa Postal 68528, Rio de Janeiro, Brazil}


\begin{abstract}
  In the   holographic AdS/QCD approach,  the  confinement/deconfinement transition is associated with the Hawking-Page transition of a thermal anti-de Sitter (AdS)  space to  an AdS  black hole. 
 In the case of the hard wall model,  the thermal transition takes place in the planar AdS thanks to the  introduction of an infrared cut-off in the geometry. 
 The corresponding thermodynamic entropy of the  $SU(N) $ gauge theory jumps from proportional to $N^0$  in the confined hadronic phase to proportional to   $N^2$ in the plasma phase, corresponding to the presence of the color degrees of freedom. The Hawking-Page transition is understood  by considering a semiclassical picture of a system consisting of two different  geometries that are asymptotically AdS. One is the AdS black hole and the other the thermal AdS space. The relative stability between these competing geometries varies with the temperature.  So, the transition is essentially a problem of stability. An interesting tool to study stability of physical systems is the configuration entropy (CE), inspired in the  Shannon informational entropy. 
 In this work we investigate the CE for the case of the AdS/QCD hard wall model at finite temperature.  
  We propose a regularized form for the    energy densities of the  black hole (BH) and of the thermal AdS geometries that makes it possible to calculate their  CEs as a function of the temperature. We find a relation between stability and the value of the CE for the system of asymptotically AdS geometries. Remarkably, it is found that the CE is proportional to $\log(T)$, where $T$ is the temperature. This result makes it possible to write out a simple relation between the configuration and the thermodynamic  entropies.          
\end{abstract}

\maketitle
 

\section{Introduction}
Soon after the AdS/CFT correspondence was proposed \cite{Maldacena:1997re,Gubser:1998bc,Witten:1998qj}, it was pointed out  by Witten \cite{Witten:1998zw} that one could associate the  Hawking-Page (HP) transition \cite{Hawking:1982dh},  in the gravitational side  of the correspondence, with the confinement/deconfinement transition in the gauge theory side.  
The HP transition emerges from a semiclassical analysis of a system consisting of  geometries that have constant negative curvature and are asymptotically anti-de Sitter on the boundary. The amplitudes for the prevalence of each of  the two ``competing'' geometries: anti-de Sitter black hole (AdS-BH) and Thermal AdS (ThAdS), depends on $\exp  ( - S) $, where $S$ is the Euclidean action integral. For the case of spherical boundary/spherical horizon there is a finite temperature $T_c$ where the action integrals of   AdS-BH   and ThAdS spaces coincide. For larger temperatures the black hole action is smaller and thus this geometry dominates or, in other words, is stable. For $ T < T_c$ the thermal AdS is the stable geometry. 

For geometries with a planar boundary there is no HP transition because the action integral of the black hole is smaller than the  one of the ThAdS, except at  zero temperature,  when the geometries coincide \cite{Witten:1998zw}.  On the other hand,  it is possible to find the  HP  transition in asymptotically AdS spaces having a planar boundary, that are relevant for studying QCD-like gauge theories, if one follows a phenomenological approach to gauge gravity duality. In the so-called AdS/QCD bottom up framework,  one gives up satisfying Einstein equations and introduces a modification in the geometry,  corresponding to adding some energy parameter of the gauge theory. In particular, the hard wall model \cite{Polchinski:2001tt,BoschiFilho:2002ta,BoschiFilho:2002vd},  makes it possible to calculate hadronic masses by assuming an approximate duality between gravity in AdS space with a hard infrared cutoff in the geometry and a non conformal gauge theory on the planar boundary.  It was shown in \cite{Herzog:2006ra,BallonBayona:2007vp}  that the hard wall model presents a HP transition at some finite critical temperature $T_c$. Interestingly, the thermodynamic entropy for the gauge theory is proportional to $N^0$ for $ T < T_c$ and to $ N^2 T^3$ for $ T > T_C$,  with $N$ the color group index,  showing an increase of degrees of freedom, associated with deconfinement of color. 

In summary, the AdS/QCD hard wall model represents the confinement/deconfinement transition of a gauge theory in terms of a change in the stability condition of a black hole geometry.   On the other hand, in the recent years,  it was show that the configuration entropy  (CE) \cite{Gleiser:2011di,Gleiser:2012tu,Gleiser:2013mga}  plays an important role in the analysis of  stability of many different physical systems, as for example in 
        \cite{Bernardini:2016qit,Braga:2017fsb,Braga:2018fyc,Bernardini:2018uuy,Ferreira:2019inu,Braga:2019jqg, Correa:2015vka,Braga:2016wzx,Karapetyan:2016fai,Karapetyan:2017edu,Karapetyan:2018oye,Karapetyan:2018yhm,Lee:2018zmp,Bazeia:2018uyg,Ma:2018wtw,Zhao:2019xle,Ferreira:2019nkz,Braga:2019jqg,Bernardini:2019stn,Karapetyan:2019ran,Braga:2020myi,Karapetyan:2020yhs,Ferreira:2020iry,Alves:2020cmr,MarinhoRodrigues:2020ssq}. Comparing the CE of the states of many different  physical systems, it was observed that, the smaller is the value of the CE, the more stable is the state. The purpose of the present work 
is to understand if the same behavior appears in the case of the transition from confining to deconfining geometries in the AdS/QCD hard wall model. Following this idea, we will develop an approach for calculating the configuration entropy for the  AdS-BH  and  ThAdS spaces and analyse the dependence of the CE on the temperature and the corresponding stability of the associated states. 

This work is organized as follows. In section (2) we review the hard wall model at finite temperature and explain how does the  Hawking-Page transition takes place. In section (3) we develop an approach for expressing the energy density of the black hole in a regularized way, that is essential for the calculation of the CE in the black hole geometry. Section (4)  is devoted to the calculation of the CE and to the analysis of the results. A study of the relation between the thermodynamic  and the configuration entropies is presented in section (5), where  a simple relation between these quantities is found. In section (6) we explore the possibility of a relation between the results obtained in this letter and the so-called Gubser-Mitra conjecture. Our  final conclusions are in section (7) and some technical details are shown in the appendices.

\section{Holographic model for confinement/deconfinement: \break Overview}
\subsection{Hard wall AdS/QCD model}
One considers  the five-dimensional Einstein-Hilbert action with a negative cosmological constant and looks for geometries with boundary $R^3 \times S^1 $ \cite{Herzog:2006ra,BallonBayona:2007vp}.There are two solutions. The first one is  thermal AdS (ThAdS), described in  the Euclidean case  by 
\begin{equation}\label{thermalAdS}
    ds^2= \frac{L^2}{z^2}\left( dt^2 + d\overrightarrow{x}^2 + dz^2 \right)\;,
\end{equation}
\noindent where  $L$ is AdS space radius. The second solution is the AdS  black hole   ( BH-AdS)  
\begin{equation}\label{BHAdS}
ds^2= \frac{L^2}{z^2}\left( f(z) dt^2 + d\overrightarrow{x}^2 + \frac{dz^2}{f(z)} \right)\;,
\end{equation}
with $f(z) = 1 - z^4/z_h^4$, being $z_h$ the black hole horizon. In both cases, one considers a compact time coordinate. For the black hole, the time has a period $\beta$ and the temperature is $T = 1/\beta = 1/(\pi z_h)$, to avoid a conical singularity of the metric on the horizon \cite{Hawking:1982dh}.

 The on-shell action for both geometries is:
\begin{equation}
I_{\text{on-shell}} = \frac{4}{L^2\kappa^2} \int d^5x \sqrt{g}\;.
\label{Action01}
\end{equation}
where $\kappa$ is the gravitational coupling. 
The hard wall model \cite{Polchinski:2001tt,BoschiFilho:2002ta,BoschiFilho:2002vd} consists of introducing a cut-off in the geometry in the form of a maximum value  for the coordinate $z$: 
 $ \, 0 \le z \le z_0$. The inverse of  $z_0$ is interpreted as an infrared energy cut-off  in the gauge theory side.  
 
The action integral of eq. (\ref{Action01}) is singular at $ z \to 0$  for both the  ThAdS and the  BH-AdS spaces.  Defining an action density $\mathcal{E} = I/V$, being $V$ the trivial spacial volume $\int d^3x$ over the components $\overrightarrow{x}$, the regularized action density $\mathcal{E}(\epsilon)$ for the ThAdS is defined as
\begin{equation}\label{EAdS}
 \mathcal{E}_{AdS}(\epsilon) = \frac{4L^3}{\kappa^2} \int_0^{\beta^\prime}  dt \int_\epsilon^{z_0} dz\, z^{-5} = \frac{L^3}{\kappa^2}\beta^\prime\left(\frac{1}{\epsilon^4} - \frac{1}{z_0^4} \right)\;,
\end{equation}
where $\epsilon$ is an ultraviolet regulator. Analogously, for the BH-AdS one has 
\begin{equation}\label{EBH}
 \mathcal{E}_{BH}(\epsilon) = \frac{4L^3}{\kappa^2} \int_0^{\pi z_h}  dt \int_\epsilon^{\text{min}(z_0, z_h)} dz\, z^{-5} = \frac{L^3}{\kappa^2}\beta\left(\frac{1}{\epsilon^4} - \frac{1}{\bar{z}^4} \right)\;,
\end{equation}
with $\bar{z} \equiv \text{min}(z_0, z_h)$. Requiring that the two geometries have the same asymptotic form  at $z = \epsilon \to 0 $, so that the time periodicity in the limit $\epsilon \rightarrow 0$ must be the same, one finds the condition $\beta^\prime = \pi z_h \sqrt{f(\epsilon)}$ \cite{Herzog:2006ra}. Using this expression for $\beta^\prime$, the difference between the actions densities, defined as
\begin{equation}
\bigtriangleup\mathcal{E} = \lim_{\varepsilon \rightarrow 0}{\left[\mathcal{E}_{BH}(\varepsilon) - \mathcal{E}_{AdS}(\varepsilon)\right]}\;,
\end{equation}
is independent of $\epsilon$ 
\begin{equation}\label{HPtransition}
\bigtriangleup \mathcal{E} = \begin{cases} 
\frac{L^3 \pi z_h}{\kappa^2}\frac{1}{2 z_h^4}\;, & \mbox{if } z_0<z_h\; , \\ \frac{L^3 \pi z_h}{\kappa^2}\left(\frac{1}{z_0^4}-\frac{1}{2 z_h^4}\right)\;, & \mbox{if } z_h<z_0 \;.
\end{cases}
\end{equation}
One notices that the  actions are equal at the critical temperature:  
\begin{eqnarray}\label{Tc}
T_c = 2^{1/4}/(\pi z_0)\;.
\end{eqnarray}
This corresponds to the Hawking-Page transition temperature.  
If $T < T_c$, the thermal AdS state dominates as $\bigtriangleup \mathcal{E} > 0$, and the black hole is unstable.  When $T > T_c$, $\,\,\bigtriangleup \mathcal{E}$ is negative, and the black hole becomes the stable geometry. This is    the holographic description of confinement/deconfinement transition developed in \cite{Herzog:2006ra}.  
  We will see now that  it is possible to obtain finite actions  in the $\epsilon \to 0 $ limit  for each space using holographic renormalization.    
  

\subsection{Holographic renormalization and thermodynamics}
In this procedure \cite{Balasubramanian:1999re,Emparan:1999pm}, the  ultraviolet divergences  are removed by adding to the action a surface counterterm. In ref. \cite{BallonBayona:2007vp},  it was pointed out that besides the usual volumetric action of eq. (\ref{Action01}),  in order to describe the planar AdS geometries considered here,  one needs also to include in the action the Gibbons-Hawking surface term. This type of boundary  term comes from the variational principle for a gravity theory with a boundary \cite{Gibbons:1976ue}. In the present case it reads
 \begin{eqnarray}
I_{GH}  =  - \frac{1}{\kappa^2} \int_{\partial M} d^4x \sqrt{h}\,K\;,
\label{GH}
\end{eqnarray}
where $K$ is the trace of the extrinsic curvature of the boundary.   The details of the calculation of the  Gibbons-Hawking surface term can be found in \cite{BallonBayona:2007vp}.  In short, at a boundary hypersurface, $    K =  \frac{1}{\sqrt{g}} \partial_a ( \sqrt{g}\, n^a )\,,$ where $n^a$ is a unitary vector normal to the boundary, see \cite{Arutyunov:1998ve}. One gets the following surface terms,  divided by the trivial spacial volume factor: 
\begin{eqnarray}
\mathcal{E}^{GH}_{AdS} &=& - \frac{4 L^3}{\kappa^2} \frac{\beta^\prime}{\epsilon^4}\;, \\
\mathcal{E}^{GH}_{BH} &=& - \frac{4 L^3}{\kappa^2} \beta\left (\frac{1}{\epsilon^4}- \frac{1}{2z_h^4}\right)\;,
\end{eqnarray}

Adding these surface terms to the corresponding volumetric ones, given in eqs. (\ref{EAdS})  and
(\ref{EBH}) one finds it out that in order to cancel out  the total  divergencies one should take   the counterterm action, for both geometries, as
\begin{equation}\label{Ictfinal}
    I_{ct} = \frac{1}{\kappa^2} \int_{\partial M} d^4 x \sqrt{h}\, \frac{3}{L}\;,
\end{equation}
where $h$ the determinant of the boundary induced metric $h_{\mu\nu}$.The holographic renormalization procedure is concluded  by  defining total actions for the geometries as $I_{total} =  I + I_{GH} + I_{ct} $.
The corresponding  action densities are given by
\begin{eqnarray}
\mathcal{E}^{total}_{AdS} &=& - \frac{L^3}{\kappa^2} \frac{\beta^\prime}{z_0^4}\;, \label{EAdShr}\\
\mathcal{E}^{total}_{BH} &=& - \frac{L^3}{\kappa^2} \beta\left( \frac{1}{\bar{z}^4}- \frac{1}{2z_h^4}\right)\;.\label{EBHhr}
\end{eqnarray}
  So, using the holographic renormalization procedure, it is possible to define finite actions for both ThAdS and BH-AdS, not only for their difference.   

From the finite actions \eqref{EAdShr} and \eqref{EBHhr} one can determine the thermodynamic entropy for each geometrical phase, and use it to characterize the confinement/deconfinement phase transition. In the saddle point approximation, using $\kappa^2 = 8\pi G_5$, and the relation between $G_5$ and the fundamental string scale, see \cite{BallonBayona:2007vp} for details, the associated thermodynamic entropies from the expression 
\begin{equation}
S = \beta \langle E \rangle + \log Z \approx \beta \langle E \rangle - I_{total}\;,\label{Entropydensity}
\end{equation}
where $\langle E \rangle =  - \frac{\partial}{\partial \beta} \log Z \approx  \frac{\partial}{\partial \beta} I_{total}$ is the expectation value of the energy, yields
\begin{eqnarray}\label{entropyAdS}
S_{AdS} &=& 0 \;, \quad \,\,\,\,\,\,\,\,\,\,\,\,\,\,\,\,\,\,\,\, \text{if} \quad T < T_c\;, \\
S_{BH} &=&  \frac{ N^2 \pi^2}{2} T^3\;, \quad\, \text{if} \quad T > T_c\;.\label{entropyBH}
\end{eqnarray}
Where there is a  jump from $N^0$ to $N^2$ dependence of  the entropy, representing the change from the confined  phase to the deconfined one \cite{BallonBayona:2007vp}, when the color degrees of freedom are free. Such a result is consistent with the free energy of the $\mathcal{N}=4$ super Yang-Mills theory at the strong coupling limit \cite{Gubser:1996de,Gubser:1998nz}, and shows  that one can use the thermodynamic entropy to identify the confinement/deconfinement phase transition.

\section{ Thermal AdS and BH AdS masses and regularized densities }
In order to compute the configurational entropy (CE)  from the finite actions \eqref{EAdShr} and \eqref{EBHhr}, one needs to calculate the associated energy (or mass) densities.  For both cases, the mass can be determined from the energy expression \cite{Hawking:1982dh, Witten:1998zw} 
\begin{equation}\label{M}
    E = \frac{\partial I}{\partial \beta} = M\;,
\end{equation}
in natural units. The results obtained in the last section, in particular \eqref{HPtransition}  shows that  the black hole is stable in the region $ z_h \leq 2^{-1/4} z_0$. Thus, one can take $\bar{z} = \text{min}(z_h, z_0)$ as $z_h$ in  expression \eqref{EBHhr}. 

Hence, applying \eqref{M} to the regularized actions constructed from the holographic renormalzation, one finds the masses of the spaces, in the range of temperatures where they are stable:
\begin{eqnarray}
M_{AdS} &=& - \frac{L^3}{\kappa^2} \frac{1}{z_0^4}\;, \,\,\,\,\,\,\,\,\quad \text{for} \quad z_h > 2^{-1/4} z_0\;, \label{MAdS} \\
M_{BH} &=& + \frac{3L^3}{2\pi^4 \kappa^2} \frac{1}{z_h^4}\;, \quad \text{for } \quad z_h < 2^{-1/4} z_0\;,.\label{MBH}
\end{eqnarray}
This means  that $M_{BH} \sim 1/z_h^4$, while $M_{AdS} \sim 1/z_0^4$. The negative sign of the ThAdS mass is interpreted as a consequence of the subtraction of the region $ z >  z_o $ in the  hard wall model. If one takes the limit $ z_0 \to \infty $ the mass goes to zero, corresponding to the mass of the empty AdS space. 


The CE  is defined as a function of the energy density  that should, in principle, be  related to the total mass  by
 \begin{eqnarray}\label{naiveintdensity}
   \int_0^{z_f} dz\, \rho(z)  = M \;, \quad \text{with} \quad z_f \equiv \{z_h, z_0\}\;,
\end{eqnarray} 
with the two possible values of the upper integration limit corresponding,  respectively, to  the BH and AdS cases. 

 It is reasonable  to assume that the mass density  should not dependent on the infrared cut-off  $z_0$ since this parameter is not part of the original AdS geometry. It is just an energy scale introduced in order to make an effective description of  a QCD like theory. So, in order to find a mass with a dependence on $  1/z_f^4$,  searching for a power series solution, one would find the simple form
 \begin{equation}\label{nonregrho}
 \rho(z) = \frac{C}{z^5} \;.   
\end{equation}
 with the constant $C$  taking different values for the AdS and BH cases. The problem one faces using such a density is that one gets the mass but also a singular term coming from the $ z \to 0 $ limit of the integral (\ref{naiveintdensity}).  In order to avoid such singularities, 
 one needs to find out  a consistent regularization process. An interesting approach is to define regularized  densities  $\rho(z)$,   related to the total mass by
 \begin{eqnarray}\label{intdensity}
 \lim_{\epsilon \to 0 }  \int_\epsilon^{z_f} dz\, \rho(z) \sim \frac{1}{z_f^4}\;, \quad \text{with} \quad z_f \equiv \{z_h, z_0\}\;.
\end{eqnarray} 
 As we shall see, such a regularization can be implemented in a similar way for both geometries. 
 We propose the following expression for the regularized black hole mass density,
\begin{equation}\label{regrhoBH}
 \rho_{BH}(z) = -\lim_{\epsilon \rightarrow 0} \frac{6 L^3}{\pi^4\kappa^2}\frac{1}{z^5}\cos(\frac{2\pi \epsilon^4}{z^4})\;.   
\end{equation}
One can easily verify that for small $\epsilon $,
\begin{eqnarray}\label{intregrho}
\int_\epsilon^{z_h}\rho_{BH}(z)\,dz &=&  \frac{3L^3}{2\pi^4\kappa^2} \frac{1}{2\pi \epsilon^4} \sin(\frac{2\pi\epsilon^4}{z_h^4}) \nonumber \\ 
&=& M_{BH} + \mathcal{O}(\epsilon^8)\;.
\end{eqnarray}
So, taking the $\epsilon \to 0 $ limit after the spatial integration,  one obtains the black hole mass.  
 
The regularized density for the thermal AdS mass \eqref{MAdS} can be obtained from the same expression constructed in the regularization of the black hole. The only difference is the constant factor, namely,  
\begin{equation}\label{regrhoAdS}
 \rho_{AdS}(z) = +\lim_{\epsilon \rightarrow 0} \frac{4 L^3}{\kappa^2}\frac{1}{z^5}\cos(\frac{2\pi \epsilon^4}{z^4})\;.   
\end{equation}
Using this density in eq. (\ref{intdensity}) one finds the ThAdS mass. An important issue, that will be addressed in the next sections is that our results must be independent of $ \epsilon $. Or, in other words, the limit $ \epsilon \to 0 $ must be well defined.

\section{Configurational entropy and Hawking-Page transition}
The configuration entropy  (CE) \cite{Gleiser:2011di,Gleiser:2012tu,Gleiser:2013mga} is inspired in the Shannon information entropy \cite{shannon}, that is a measure of  information content. 
For a variable  that can take  $N_d$ discrete possible values, with probabilities given by $p_i$, it is defined by
\begin{equation}\label{Shannon}
S_{\text{info}}=- \sum_{i=1}^{N_d} p_i \ln p_i\;.
\end{equation}  
The configurational entropy for a physical system is defined as a continuous version of (\ref{Shannon}). It is defined in terms of the modal fraction, which is constructed upon of the Fourier transform of the energy (mass) density, $\widetilde{\rho}(k)$, that describes the corresponding physical states. For the black hole and thermal AdS states in the hard wall model, the energy density is  a function of the Poincaré coordinate $z$, according to \eqref{regrhoBH} and \eqref{regrhoAdS}, so that
\begin{equation}
 \widetilde{\rho}(k) = \frac{1}{2\pi}  \int dz\, \rho(z) e^{i k z}\;.
\end{equation} 
The modal fraction is defined as 
\begin{equation}\label{modal}
f(k) = \frac{ \vert \widetilde{\rho}(k) \vert^2 }{\mathcal{N}}\;,
\end{equation}
where the normalization constant is defined as:
\begin{eqnarray}\label{normalization}
\mathcal{N} = \int dk \, \langle \vert \widetilde{\rho}(k) \vert^2 \rangle\;, 
\end{eqnarray}
The corresponding configurational entropy for such localized energy densities  is then defined as the functional \cite{Gleiser:2012tu}
 \begin{equation}\label{CE}
S_C[f] = - \int dk\, f(k) \ln f(k) \;.
\end{equation}
In the Appendix B we show that  $  - f(k) \ln f(k)  $ is  always positive. From the Fourier transforms of $\rho_{BH}(z)$ and $\rho_{AdS}(z)$,
\begin{eqnarray}
 \widetilde{\rho}_{BH}(k) &=& \frac{1}{2\pi}  \lim_{\epsilon \to 0}  \int_{\epsilon}^{z_h} dz\, \rho_{BH}(z) e^{i k z}\;,\label{FrhoBH}\\
 \widetilde{\rho}_{AdS}(k) &=& \frac{1}{2\pi}  \lim_{\epsilon \to 0}   \int_{\epsilon}^{z_0} dz\, \rho_{AdS}(z) e^{i k z}\;,\label{FrhoAdS}
\end{eqnarray}
one can compute the modal fractions for the thermal and black hole AdS spaces using \eqref{modal}, and then obtain the configurational entropies as functions of the temperature, above and bellow $T_c$. From this, one can finally study the confinement/deconfinement phase transition from the point of view of stability. Replacing \eqref{regrhoBH} and \eqref{regrhoAdS} into \eqref{FrhoBH} and \eqref{FrhoAdS}, respectively, one finds out  that $\widetilde{\rho}_{BH}(k)$ and $\widetilde{\rho}_{AdS}(k)$ do not possess analytical solutions. So,  the CE is calculated using numerical methods.   


The numerical computation of the Fourier transform of the energy density  in eqs. (\ref{FrhoBH}) 
and  (\ref{FrhoBH})  is performed with a finite value of the UV regulator  $ \epsilon $. The limit 
$ \epsilon  \to 0 $  is obtained,  in the numerical approach, by identifying the order of magnitude of  vaues of $ \epsilon $ for wich  taking smaller values would  not change the results.  We found it out that  from $\epsilon/z_o \sim  10^{-13}$ to smaller values there is no change in any of the results of this work. So, this value of $\epsilon  $ was used in our computations. 
 In the Fourier space, the squared absolute value of the black hole energy density, that defines the normalization factor and the modal fraction, can be written as 
\begin{equation}
    \vert \widetilde{\rho}_{BH}(k) \vert^2 = \left[  \lim_{\epsilon \to 0} \frac{1}{2\pi} \int_{\epsilon}^{z_h} dz\,  \rho_{BH}(z) \cos(kz)\right]^2 + \left[  \lim_{\epsilon \to 0}  \frac{1}{2\pi}\int_{\epsilon}^{z_h} dz\, \rho_{BH}(z) \sin(k z)\right]^2 \;.
\end{equation}
For the the thermal AdS, $\vert \widetilde{\rho}_{AdS}(k) \vert^2$ is similar, only replacing $\rho_{BH}(z) \rightarrow \rho_{AdS}(z)$ in the expression above. 

With all these information we are now able to compute the modal fractions for each space, and finally determine the corresponding configurational entropies in the confined and deconfined phases. The transition of geometries occurs at $\frac{\beta_c }{\pi z_0} \, = \frac{1}{T_c \pi z_0}  =  0.840896$. Above $T_c$ the space is AdS BH and below $T_c$ it is the ThAdS. We take $z_0 = 1$ and show in table of TABLE. 1 the values of the BH CE for different values of $ \beta/\pi $, which is represented by the points in FIG. 1 for $\beta/\pi < \beta_c/\pi$. 
\begin{table}[h]
\centering
\begin{tabular}[c]{|c|c||c|c|}
\hline 
 $\beta/\pi z_0 $ &  Black hole CE  & $\beta/\pi z_0 $ & Black hole CE   \\
\hline
$\,\,\,0.025\,\,\,$ & $ 17.5229$ & $\,\,\,0.4\,\,\,$ & $ 14.7761 $ \\
\hline
$\,\,\,0.038\,\,\,$ & $ 17.1049 $    & $\,\,\, 0.45\,\,\,$ & $14.6326 $ \\
\hline 
$\,\,\,0.05\,\,\,$ & $  16.8300 $   & $\,\,\,0.5\,\,\,$ & $ 14.5268 $ \\
\hline
$\,\,\, 0.075\,\,\,$ & $ 16.4244  $ & $\,\,\,0.55\,\,\,$ & $   14.4321 $    \\
\hline
$\,\,\,0.1\,\,\,$ & $ 16.1424  $ & $\,\,\,0.6\,\,\,$ & $ 14.3455 $   \\ 
\hline
$\,\,\,0.15\,\,\,$ & $ 15.7313 $    & $\,\,\,0.65\,\,\,$ & $ 14.2650 $ \\
\hline 
$\,\,\,0.2\,\,\,$ & $ 15.4437 $   & $\,\,\,0.7\,\,\,$ & $14.1940  $ \\
\hline
$\,\,\,0.25\,\,\,$ & $15.2204 $ &  $\,\,\,0.75\,\,\,$ & $  14.1219 $    \\
\hline
$\,\,\,0.3\,\,\,$ & $ 15.0380  $ & $\,\,\,0.8\,\,\,$ & $  14.0573 $   \\ 
\hline
$\,\,\,0.35\,\,\,$ & $  14.8840 $    & $ 0.840896$ & $14.0074 $ \\
\hline 
\end{tabular}   
\caption{Black hole configurational entropies at different temperatures.}
\label{table1}
\end{table}

\begin{figure}[!htb]\label{fig4}
	\centering
	\includegraphics[scale=0.52]{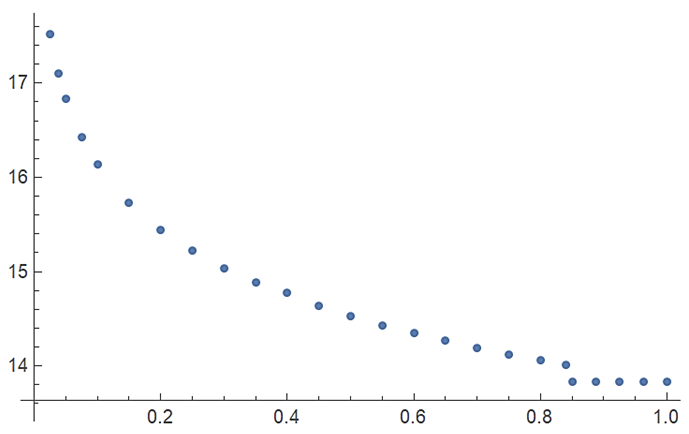}
	\caption{CE versus $\beta /\pi z_0 $.}
\end{figure}

For temperatures below $T_c$, see FIG. 1 for $\beta/\pi z_0 > 0.840896$, the CE is constant, given by 
\begin{equation}
S_C^{AdS} =  13.8344\;,
\end{equation} 
indicating that the ThAdS does not change the stability condition up to the critical temperature, where the BH is in the stable phase. Above $T_c$ the CE depends on the temperature according to the description of FIG. 1. As we increase the temperature, \textit{i.e.}, as we decrease $\beta$, the CE increases.  From the usual behavior of the CE, this would mean that black holes are more unstable at higher temperatures. This is consistent with the fact that  black holes radiate
and the radiation effect becomes stronger at higher temperatures, leading to a loss of energy.


\section{Relation between the thermodynamic and configuration entropies}
The fact that we calculated the configuration entropy for a system that is in thermal equilibrium and therefore has a well defined thermodynamic entropy  opens up the possibility of answering an interesting question. We mean: is there a simple  relation between these two quantities? In this section we want to investigate this possibility.
We now write the temperature in units of $ 1/z_0$ or,  in other words, the dimensionless temperature that we use from now on is equal to the dimensionful temperature multiplied by the energy parameter $z_0$ of the hard wall model. 
In order to compare the black hole thermodynamic  entropy, which is proportional to $T^3$, with the CE  one must find out  its temperature dependence. By plotting CE versus temperature for $T>T_c$, one obtains the points of Figure 2-(A), corresponding to the CE values displayed in the TABLE 1. This plot suggests a logarithmic behavior. In order to see if this is true, we plot in Figure 2-(B) the black hole CE versus $\log T$, which is given by a straight line. See TABLE 2 in Appendix A for the values of $T$ and $\log(T)$ corresponding to each CE of TABLE 1, which were used to plot the points of FIG 2-(A) and (B).  From the Figure 2-(B) we conclude that the BH configurational entropy is proportional to $\log T$, 
\begin{equation}\label{BHCElogT}
    S_C^{BH}(T) = A_0 \log T + B_0\;, 
\end{equation}
being $A_0$ and $B_0$ constants, that can be numerically estimated as $A_0 = 0.99 \pm 0.02$ and $B_0 = 14.99 \pm 0.02$. 
\begin{figure}[!htb]\label{fig4}
	\centering
	\includegraphics[scale=0.52]{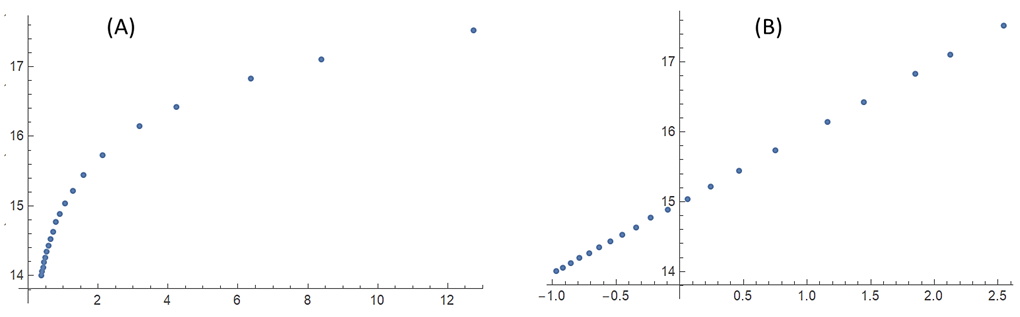}
	\caption{(A) Black hole CE versus $T$; (B) Black hole CE versus $\log T$.}
\end{figure}
 
 Now, using eq. \eqref{BHCElogT}, we can relate the configurational entropy to the thermodynamic one. This can be done rewriting the constants $A_0$ and $B_0$ in the form:  
\begin{equation}\label{redefinition}
    A_0 = 3 A\;, \quad \text{and} \quad  B_0= B+A\log\left(\frac{ N^2 \pi^2}{2}\right)\;,  
\end{equation}
so that, 
\begin{equation}\label{relation}
    S_{BH} = e^{-\frac{B}{A }} \exp\{ S_C / A \}\;, 
\end{equation}
  Equation \eqref{relation} shows the relation between the thermodynamic BH entropy $S_{BH}$ with the configurational one $S_C$, being the first given by the exponential of the second up to constant factors.  So, we found a simple relation between these two quantities. 
For the confined phase the situation is trivial:  the thermodynamic entropy vanishes and  the CE is constant.  
 
 \section{Comparison with the Gubser-Mitra conjecture}
 
  It is interesting to check if the results for stability of the hard wall model analyzed in the previous sections are consistent with the so-called Gubser-Mitra (GM) conjecture \cite{Gubser:2000mm,Gubser:2000ec}, that relates dynamical stability of a black hole with infinite horizon area to the thermodynamic stability. The Hawking Page  transition, considered in this letter,  is obtained from a semiclassical analysis of thermodynamic stability (see \cite{Monteiro:2010cq} for a  review). It corresponds to  a global phase transition between two different geometries that are asymptotically AdS spacetimes with compact Euclidean time coordinate.   The result that emerges from the analysis of the Hawking-Page transition for the hard wall model is that for $ T > T_c$ the black hole space is thermodynamically stable, while for $T < T_c$ the thermal AdS space is thermodynamically stable.  
  
 Regarding dynamic stability, one should study the behavior of the  black hole and thermal AdS geometries under the effect of perturbations of the metric. The hard wall model is a phenomenological framework where one places a hard cut-off in the geometry in order to obtain results of hadronic physics assuming gauge/gravity duality to hold. The hard wall geometry, with or without a black hole, is not a solution of the classical equations of motion (the Einstein equations). So, it would make no sense to investigate perturbations in the equations of motion starting from such geometries. Nevertheless, there is an interesting way of making contact with the GM conjecture. 
 As discussed in \cite{Gubser:2000mm}, an intuitive explanation for the dynamical instability of a black hole of infinite extent  is that the entropy of an array of finite size black holes is higher  than the entropy of the infinite size uniform black hole with the same mass. In other words, dynamical instability would appear as a consequence of the possibility of a transition to other geometry with larger entropy. We presented in equations (\ref{entropyAdS}) and (\ref{entropyBH})  the entropy densities  for thermal AdS and for the black hole in the interval of temperatures where they are the dominant phase. For the thermal AdS, eq. (\ref{entropyAdS}) is valid at any temperature since the total action density of eq. (\ref{EAdShr}) does not change with $T$. But for the black hole geometry the action density of eq. (\ref{EBHhr})  depends on the relation between $z_0$ and $z_h$. The expression shown in eq.  (\ref{entropyBH}) is valid  for $ z_h  >  z_0 $ or, equivalently,  $ T > 1/(\pi z_0) = T_c/2^{1/4}$.  Calculating the entropy (per unity of transverse area) for the black hole action for $ T  < 1/(\pi z_0) $, using eq. (\ref{Entropydensity}),  one would obtain the awkward result 
\begin{equation}
S_{BH}  = - \frac{  N^2 \pi^2 T^3  }{2} \quad \text{if} \quad  T < \frac{T_c}{2^{1/4}}\; .\label{entropy2BH}
\end{equation}
 Such an unphysical negative entropy shows up because the imposition of a hard cut-off at a position $ z = z_0 < z_h $ eliminates  from the space a region containing the horizon. In this case one would have $ S_{BH}  <  S_{AdS} $   for $ T < T_c/2^{1/4}$  and $ S_{BH}  >   S_{AdS} $   for $ T > T_c/2^{1/4}$. So, for $ T < T_c/2^{1/4}$  and for $ T > T_c $ there is an agreement between thermodynamic instability and the existence of other space with larger entropy, that could be associated with dynamic instability. Only in the region  $   T_c/2^{1/4} < T < T_c$  the thermodynamically stable geometry has a smaller entropy. One should not take this analysis as a  proof of a relation between dynamical and thermodynamic stabilities for the hard wall model since this space it is not a solution of Einstein equation and also the geometry is not analytic at  $ z = z_0 $.  It is just an analyisis of self consistency of the model.  This discussion provides  a motivation for performing a similar analysis in some other  AdS/QCD model   that is consistent with Einstein equations and does not present any non analyticity. We plan to do this in a future work.

\section{Conclusions}

In this  work we calculate the configuration entropy for the two geometries that represent, within the hard wall AdS/QCD model, the two phases of a QCD-like gauge theory. One is the thermal AdS space that represents the confined phase and is dominant for temperatures below $T_c$. The other is the AdS black hole, which represents the deconfined phase that dominates for higher temperatures. We found that for the ThAdS space, the CE is constant. This is consistent with the fact that, below $T_c$ this space is stable. For the BH-AdS geometry, we found that the CE increases with the temperature. This can be seen from FIG. 1 where $ \beta $ is the inverse of the temperature, so $ \beta \to 0 $ is equivalent to $  T \to \infty $,  or in FIG. 3 where we observe that the black hole CE is proportional to $\log(T)$. Increasing CE means in general increasing instability. The reason for this can be traced to the Hawking radiation of the black hole, that increases with the temperature and causes instability in the physical state. For discussions of  Hawking radiation of AdS black holes, see for example \cite{Hubeny:2009rc}. 
So, we found it out that for the hard wall description of QCD-like thermal media, the relation between  increase in the CE and increase in the instability holds.  We also found a relation between the black hole thermodynamic and configuration entropies, being the first given by the exponential of the second, up to constant factors.  
 For an interesting study of deconfinement transition in holographic QCD using entanglement entropy see \cite{Dudal:2018ztm}.

\appendix

\section{TABLE: Black hole CE, $T$, and $\log(T)$} 
We display in this table the values of the temperature $T$ and $\log(T)$, with again $T$ in units of $1/ z_0$,  which were used to plot FIG 2-(A) and (B), for the same black hole configuration entropies of TABLE 1.
\begin{table}[h]
\centering
\begin{tabular}[c]{|c|c|c||c|c|c|}
\hline 
 $T$ & $\log(T)$ & Black hole CE  & $T$ & $\log(T)$ & Black hole CE   \\
\hline
$\,\,\,0.378537\,\,\,$ & $\,\,\,-0.971443\,\,\,$ & $14.0074 $ & $\,\,\,0.909457\,\,\,$ &$\,\,\,-0.0949078\,\,\,$ & $  14.8840 $ \\
\hline
$\,\,\,0.397887\,\,\,$ & $\,\,\,-0.921586\,\,\,$ & $ 14.0573 $ & $\,\,\,1.06103\,\,\,$ & $\,\,\,0.0592429\,\,\,$ &$ 15.0380  $ \\
\hline 
$\,\,\,0.424413\,\,\,$ &$\,\,\,-0.857048\,\,\,$ & $  14.1219 $    & $\,\,\,1.27324\,\,\,$ &$\,\,\,0.241564\,\,\,$ &$15.2204 $\\
\hline
$\,\,\,0.454728\,\,\,$ &$\,\,\,-0.788055\,\,\,$ & $14.1940 $  & $\,\,\,1.59155\,\,\,$ &$\,\,\,0.464708\,\,\,$ & $ 15.4437 $   \\
\hline
$\,\,\,0.489708\,\,\,$ &$\,\,\,-0.713947\,\,\,$ &$ 14.2650 $  & $\,\,\,2.12207\,\,\,$ &$\,\,\,0.75239\,\,\,$ & $ 15.7313 $  \\ 
\hline
$\,\,\,0.530516\,\,\,$ &$\,\,\,-0.633904\,\,\,$ &$ 14.3455 $    & $\,\,\,3.18310\,\,\,$ &$\,\,\,1.15786\,\,\,$ & $ 16.1424  $\\
\hline 
$\,\,\,0.578745\,\,\,$ &$\,\,\,-0.546893\,\,\,$ &$   14.4321 $   & $\,\,\,4.24413\,\,\,$ &$\,\,\,1.44554\,\,\,$ & $ 16.4244  $\\
\hline
$\,\,\,0.636620\,\,\,$ &$\,\,\,-0.451583\,\,\,$ &$ 14.5268 $  &  $\,\,\,6.36620\,\,\,$ &$\,\,\,1.85100\,\,\,$ &$  16.8300 $    \\
\hline
$\,\,\,0.707356\,\,\,$ &$\,\,\,-0.346222\,\,\,$ & $14.6326 $ & $\,\,\,8.37658\,\,\,$ &$\,\,\,2.12544\,\,\,$ & $ 17.1049 $   \\ 
\hline
$\,\,\,0.795775\,\,\,$ &$\,\,\,-0.228439\,\,\,$ &$ 14.7761 $    & $ \,\,\,12.73240\,\,\,$ &$\,\,\,2.54415\,\,\,$ &$ 17.5229$ \\
\hline 
\end{tabular}   
\caption{BH CE, $T$ and $\log(T)$.}
\label{table1}
\end{table}

\section{ Positivity of the CE } 
In this appendix we will discuss the positivity of the configuration entropy, defined by eq. 
(\ref{CE}). One can regard the integrand of this equation,
\begin{equation}\label{lambda}
    \lambda(k) = - f(k) \ln f(k)\;, 
\end{equation}
as representing the CE per unit of momentum. We show in Figure 3-(A) the form of  $\lambda(k) $ for the black hole case, for 3 different temperatures. One notices that in the limit $k \rightarrow 0$,  $\lambda(k)$ reaches a finite value.
\begin{figure}[!htb]\label{fig3}
	\centering
	\includegraphics[scale=0.5]{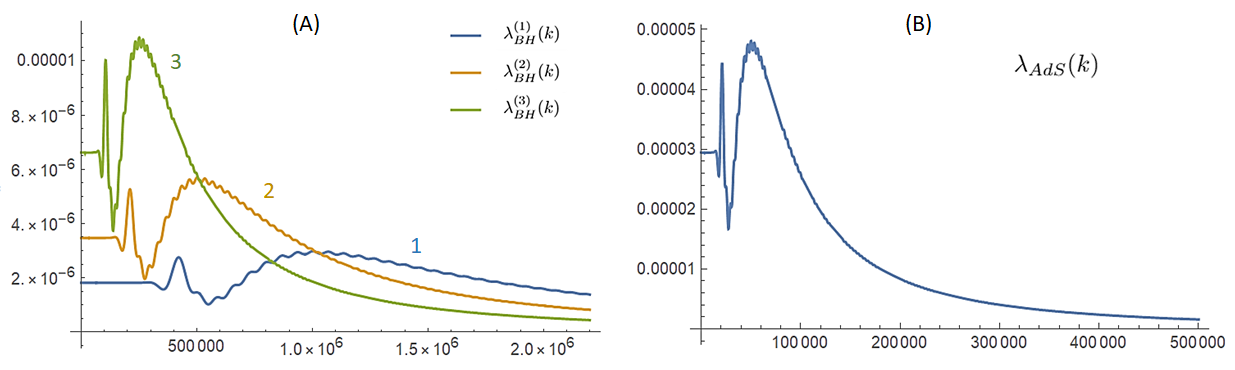}
	\caption{(A) Black hole CE per unit of $k$ versus $k$, for $T_1 = 1/(\pi 0.05)$ (blue), $T_2 = 1/(\pi 0.1)$ (orange), and $T_3 = 1/(\pi 0.2)$ (green). (B) CE per unit of momentum for the thermal AdS space.}
\end{figure}
In the black hole case, see Figure 3-(A), one sees that this value increases as the temperature increases (note that $T_1 > T_2 > T_3$). The curve, as $k$ increases, oscillates until it passes through a local maximum, then through a local minimum, where it begins to grow until reaching the global maximum, to finally tend to zero as $k \rightarrow \infty$. For higher temperatures, the global maximum is higher and is reached for lower values of momentum. In the thermal AdS space, the behavior is similar, but $\lambda(k) $ does not depend on the temperature. The corresponding plot is shown in Figure 3-(B).  
Hence, since the configurational entropy,  defined as the integration of $\lambda(k)$ over all modes, it   is always positive.  

In order to explain the physical origin of the $\lambda(k)$ behavior displayed above for different temperatures, we should take a look at the energy density in the Fourier space, see eq. \eqref{FrhoBH}. In the FIG. 4 below, we observe the same signature of the curve, whose pattern is transmitted to $\lambda^{(2)}_{BH}(k)$ of FIG 3, which depends on $\vert \widetilde{\rho}(k)\vert^2$, according to eq. \eqref{lambda} and the modal fraction definition \eqref{modal}. The small values of $\lambda(k)$, if compared to the energy density, is only attributed to the normalization factor, see eq. \eqref{normalization}. At $T_2$, for instance, $\mathcal{N}_{T_2} = 5.99696 \times 10^{57}$. As we decrease the value of the UV regulator $\epsilon$, the small oscillations along the $\lambda(k)$ curve tends to disappear. Such small oscillations must then be interpreted as a consequence of the regularization of the BH energy density employed in \eqref{regrhoBH}, these oscillations being harmless in the precision in which we are working. 
\begin{figure}[!htb]\label{fig3}
	\centering
	\includegraphics[scale=0.4]{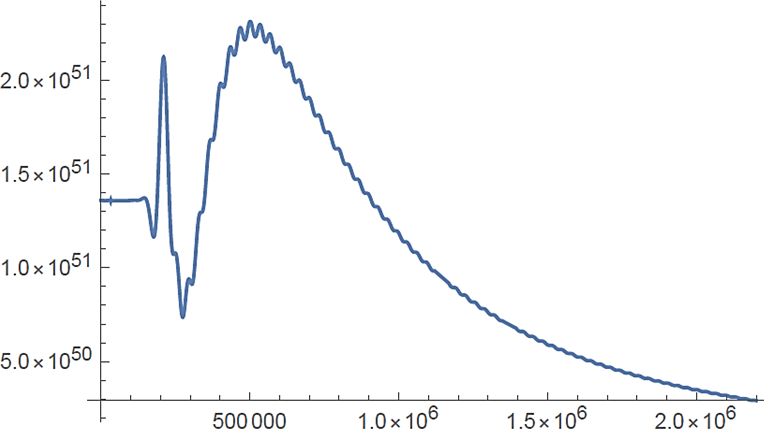}
	\caption{$\vert \widetilde{\rho}(k)\vert^2$ for a black hole at $T_2 = 1/(\pi 0.1)=3.1831$.}
\end{figure} 

One can numerically estimate the maximum values of the $\lambda(k)$ function and   the value of $ k \equiv  k_{max} $ where it occurs,  for  the temperatures analysed in FIG 3. The results are displayed in TABLE III. 
\begin{table}[h]
\centering
\begin{tabular}[c]{|c|c|c|}
\hline 
   Temperature & $ \lambda(k_{max})$ & $k_{max} $    \\
\hline
$T_1 = 6.3662  \, $  & $ \,2.9819 \times 10^{-6}\,\,\,$ &$1,00 \times 10^6/z_0$   \\
\hline
  $  T_2 =  T_1/2$ & $ \, 5.6988 \times 10^{-6}\,\,\,$ & $ 5,01 \times 10^5/z_0 $   \\
\hline
 $  T_3 =  T_1/4  $ & $\,\,\,1.0867\times 10^{-5}\,\,\,$ &  $ 2,51 \times 10^5/z_0 $   \\
\hline 
\end{tabular}   
\caption{Maximum value of $\lambda_{BH}(k)$ at different temperatures $T$ (espressed in units of $1/ z_0 $).}
\label{table1}
\end{table} 
 From this table one notices  that  $k_{max} $ is approximately proportional to  $T$.  This is consistent with the fact that the energy distribution should depend only on two dimensionful  parameters: $z_0$ and $T$.  Performing the calculations in terms of units of $1/z_0 $ we remain with only $ T $ as the free energy parameter. That is why $ k_{max}$ is proportional to $T$.

\noindent {\bf Acknowledgments:} The authors are supported by  CNPq - Conselho Nacional de Desenvolvimento Cient\'ifico e Tecnol\'ogico. This work received also support from  Coordena\c c\~ao de Aperfei\c coamento de Pessoal de N\'ivel Superior - Brasil (CAPES) - Finance Code 001.

\end{document}